# Experimental evidence that the gravitational constant varies with orientation.

by Mikhail L. Gershteyn∗†, Lev I. Gershteyn†, Arkady Gershteyn†, Oleg V. Karagioz‡

∗Massachusetts Institute of Technology, NW16-189, 167 Albany St., Cambridge, MA 02139, U.S.
† Insight Product Co., PO Box 35297, Brighton, MA 02135, U.S.
‡Tribotech division of National Institute of Aviation Technology
5-12 Pyrieva St.,Moscow 119285, Russia

**Correspondence should be addressed to Mikhail Gershteyn (MGershteyn@Aol.com).**

## Abstract

In 1687, Isaac Newton published the universal law of gravitation stating that two bodies attract each other with a force proportional to the product of their masses and the inverse square of the distance. The constant of proportionality, G, is one of the fundamental constants of nature. As the precision of measurements increased the disparity between the values of G, gathered by different groups, surprisingly increased [1-16]. This unique situation was reflected by the 1998 CODATA decision to increase the relative G uncertainty from 0.013% to 0.15 % [17]. Our repetitive measurements of the gravitational constant (G) show that G varies significantly with the orientation of the test masses relative to the system of fixed stars, as was predicted by the Attractive Universe Theory [18,19]. The distances between the test masses were in the decimeter range. We have observed that G changes with the orientation by at least 0.054%.

In 1988 two of the authors (M. L. Gershteyn and L.I. Gershteyn) published The Attractive Universe Theory, AUT, which predicted that the gravitational force between two bodies depends on the distribution of matter in the surrounding Universe. Therefore, G is a variable depending on the orientation of gravitating bodies with respect to the system of fixed stars and it could be affected by the position of close massive bodies such as the Sun. The dependence of G on direction in space has been named G anisotropy. The Attractive Universe Theory suggested that G anisotropy is the main cause of the disparity in high precision G measurements. It is necessary to note that G anisotropy on much smaller scale (0.00002%) was proposed by Will based on Whitehead's Gravitational theory [20].

In 2001, the authors of this article developed a strategy to check predicted G anisotropy by gravitational experiments conducted in Moscow under supervision of one of the authors (O.V. Karagioz). We have detected G anisotropy by two independent methods. Measurements show that the level of G anisotropy is not less than 0.054 %.

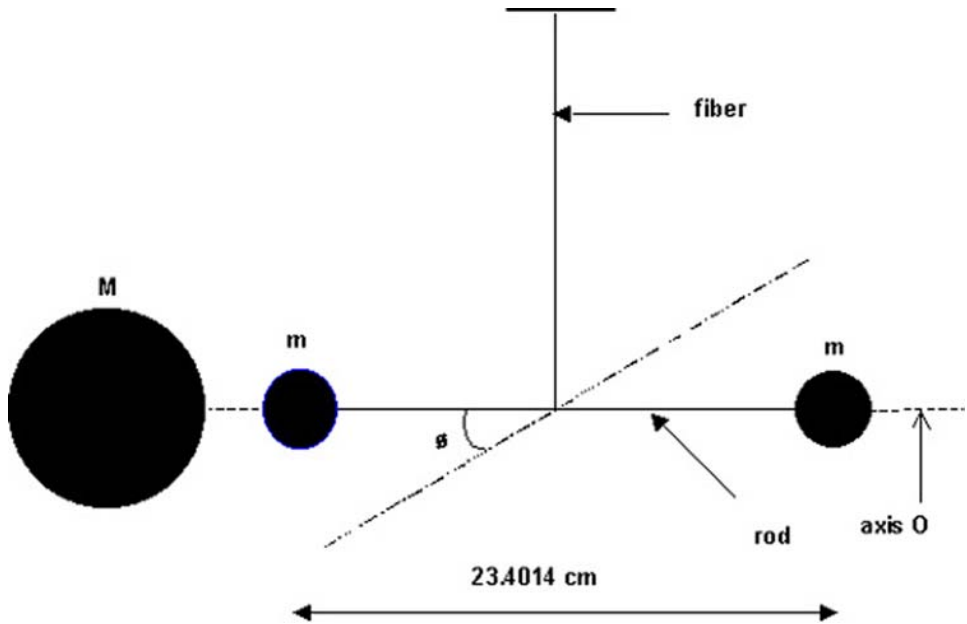

*Fig.1: Schematic of dynamic torsion balance*

The experiment used a dynamic torsion balance to measure G (Fig. 1). The dynamic torsion balance consists of two masses (m=. 942 g) connected by a rod which is suspended by a thin fiber. The torsion balance was then set into oscillatory motion in a horizontal around the equilibrium axis O. A heavy mass (M=4,287.347 g) was placed on O, near one end of the torsion balance and affected the period of oscillation.  The heavy mass (M) was moved along axis O to different distances from the torsion balance, causing the period of oscillation T to change; this change determined G .The amplitude of the angular oscillation of the torsion balance was rather small (1.6-3.1°). The entire mechanism was set on the floor. Thus the axis O rotated with the Earth. Given the above setting, G anisotropy will cause the measurements of G to change synchronically with the rotation of the Earth. If G anisotropy is connected to the system of fixed stars then the value of G must be changing with the period of one sidereal day (23.93 hr.). G was measured around the clock during a period of seven months. The time interval between two consecutive measurements of G was 1.1-1.2 hr. for most (94.8%) data points. Measurements have been performed automatically, under constant temperature (23±0.1°C) in vacuum chamber (pressure=$10^{-6}$ Pa).

The first method for identifying G anisotropy was the spectral analysis of G measurements. Lomb's "Least-squares frequency analysis of unequally spaced data" [21,22] was used in the computer program created by George B. Moody [23]. The result is a periodogram of G (Fig. 2) which has a clear absolute maximum at the period 23.89 hr.  Given the precision of the method (±0.06 hr) this period matches the sidereal day. The amplitude of the sidereal component corresponds to a 0.054% change in G with direction.

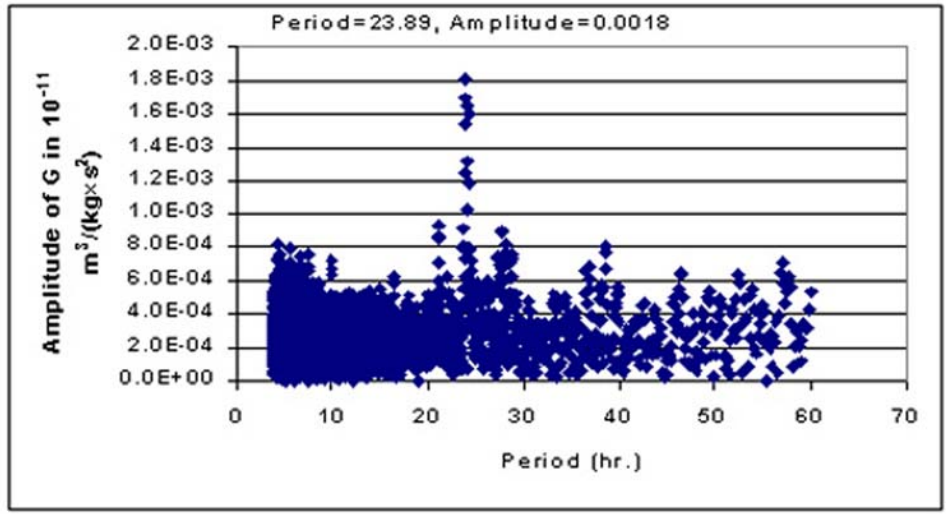

*Fig. 2: Periodogram of G*

The second method of testing G anisotropy was developed in order to demonstrate that the observed variation is in fact connected to the gravitational force between the heavy mass M and the torsion balance rather than other factors. We call the second method "the antenna for G anisotropy". The method analyzed the amplitude of sidereal component of period T for different positions of mass M. In presence of G anisotropy, the force of attraction between M and the torsion balance should change with the rotation of the Earth. The period of oscillation of the torsion balance (T) will change accordingly. The magnitude of the variation of T given M is in a certain position will depend on the force of attraction from mass M. More precisely the amplitude of T's sidereal component should be proportional to factor Z, which is the difference between one over the square of mean period of oscillation when M is in that position and when M is taken out.

Other periodical factors (tidal effects, seismic vibrations, temperature, etc.) in principal can also affect the restoring torque of the fiber and thus the period of torsion balance. In contrast to the effect of G anisotropy, any small change in restoring torque of the fiber will affect period in direct proportion to the value of period T itself. Accordingly interfering periodic signals from these factors should be directly proportional to T.

In our experiment, as M is moved into different positions, the value of T changes at a maximum of 3.2%. Thus the amplitude of the interfering signal should change by no more than 3.2% as we vary the position of M. An interfering signal should increase as the period T increases.

We conducted spectral analysis of measurements of torsion balance's period T for three different positions of the heavy mass M (Fig. 3). The results are:

1. All the periodograms have a clear absolute maximum with a period corresponding to the sidereal day given the experimental error.

2. The amplitude of "sidereal component" increases as the distance between the heavy mass M and the torsion balance decreases (Fig. 4).

3. When M is moved from the farthest position to the closest position the amplitude of sidereal component of T increased by 63%.

4. The correlation coefficient between amplitude of sidereal component versus the factor Z is 0.99.

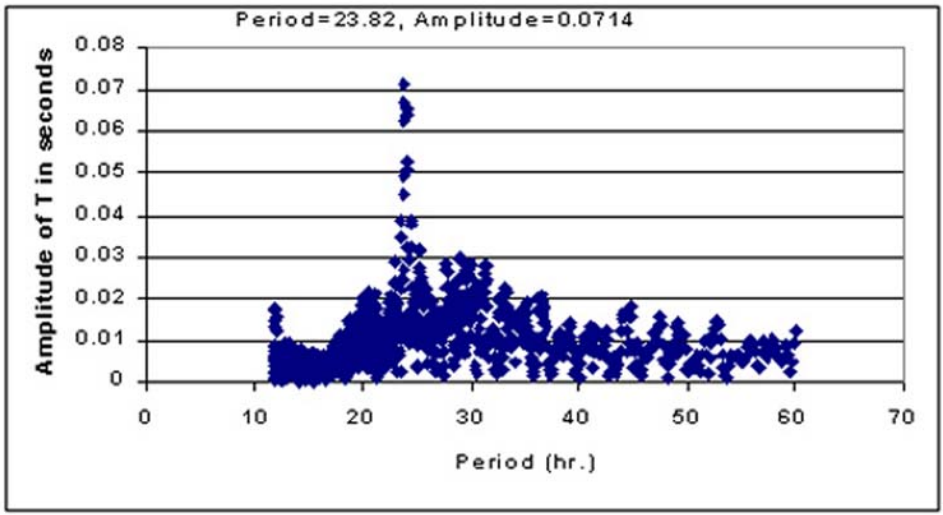

*Fig.3: Periodogram of T when M is in position 2*

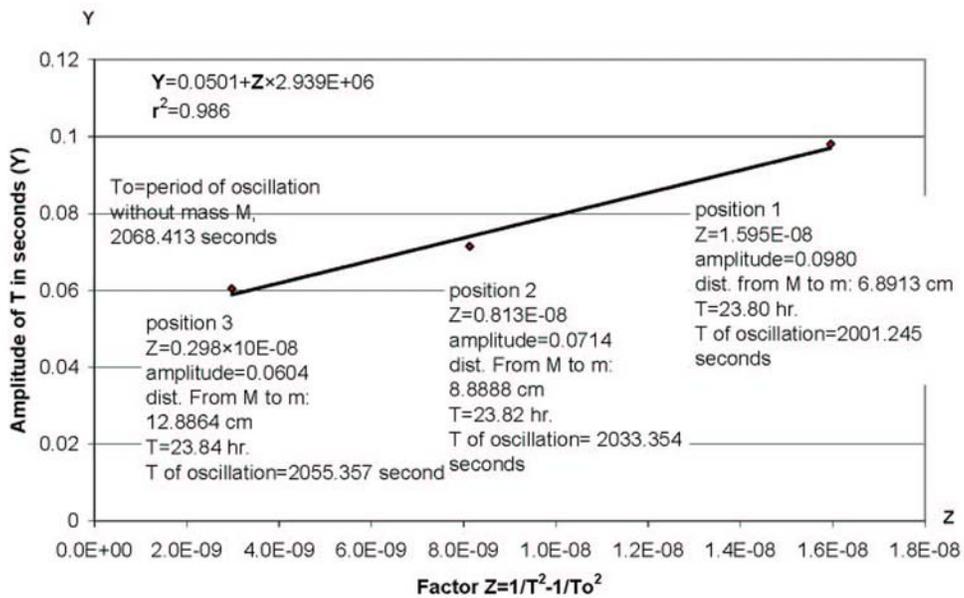

*Fig.4: Amplitude of sidereal component (Y) vs. Factor Z*

In other words, the experiment demonstrates that the torsion balance works as an antenna for detecting G anisotropy signal, with the efficiency increasing when M is moved closer to the balance. The high correlation coefficient demonstrates that the observed effect is the result of gravitational interaction rather than any other factors. The observed signal can not be caused by factors which affect a restoring torque of the fiber. As stated above, an interfering signal should decrease by a maximum of 3.2 % as T decreases; on contrary, the observed signal increases by 63% as T decreases.

The regression line in fig 4 shows that observed signal of G anisotropy contains two components: one, which drops with distance to M, and another which is a constant, presumably caused by the masses asymmetrically located in the laboratory.

The experimental setup allows us to make measurements only for a limited set of directions, so measured level of G anisotropy 0.054% is only the lower estimation for this effect. Further experiments must be performed in order to determine exact value of G in different directions relative to the stars as well as at different magnitudes of the distance between the gravitating masses.

**Aknowledgments**

We would like to express our gratitude to Lev Tsimring, Charles Welch, and Robert D. Reasenberg for their advice and valuable discussions. Also we would like to thank George B. Moody for technical assistance in utilizing his "Lomb" computer program.